# Quantitative Fluorescence Excitation Spectra of Synthetic Eumelanin

Stephen Nighswander-Rempel,* Jennifer Riesz, Joel Gilmore,
Jacques Bothma and Paul Meredith

Soft Condensed Matter Physics Group
Centre for Biophotonics and Laser Science
School of Physical Sciences, University of Queensland
St. Lucia, QLD, Australia 4072

**Abstract**
Previously reported excitation spectra for eumelanin are sparse and inconsistent. Moreover, these studies have failed to account for probe beam attenuation and emission reabsorption within the samples, making them qualitative at best. We report for the first time quantitative excitation spectra for synthetic eumelanin, acquired for a range of solution concentrations and emission wavelengths. Our data indicate that probe beam attenuation and emission reabsorption significantly affect the spectra even in low-concentration eumelanin solutions and that previously published data do not reflect the true excitation profile. We apply a correction procedure (previously applied to emission spectra) to account for these effects. Application of this procedure reconstructs the expected relationship of signal intensity with concentration, and the normalised spectra show a similarity in form to the absorption profiles. These spectra reveal valuable information regarding the photophysics and photochemistry of eumelanin. Most notably, an excitation peak at 365 nm (3.40 eV), whose position is independent of emission wavelength, is possibly attributable to a DHICA component singly linked to a polymeric structure.

**Introduction**
Melanins make up a class of pigments presiding dominantly in the skin and hair of humans, but also present in the eye, the inner ear, and the brain. There are four main categories of melanins (eumelanin, pheomelanin, allomelanin, and neuromelanin),[1,2] but only eumelanin and pheomelanin are present in human skin. Both pigments exhibit protoprotective properties derived from their high absorbance in the visible and UV, their extremely low quantum yields, and their anti-oxidant and free radical-scavenging behaviour. However, several studies have implicated these pigments along with their precursors in the development of melanoma skin cancer.[3,4] For these reasons, the photophysics, photochemistry, and photobiology of melanins are subjects of intense scientific interest.

Despite the fact that eumelanin has been widely studied for decades, several fundamental questions concerning its physics and chemistry remain unanswered. For instance, while it is well-recognized that eumelanin is an aggregate of indolic monomers, such as 5,6-dihydroxyindole (DHI) and 5,6-dihydroxyindole, 2-carboxylic acid (DHICA), it remains unclear exactly how these monomer units are connected together to form a secondary structure, even in synthetic samples.[5] The issue of how eumelanin binds with neighbouring proteins is also unsettled. Fluorescence spectroscopy has shed some light on the photophysical properties of melanins,[6,7] but despite the fact that eumelanin fluorescence was first investigated over 30 years ago, there is still some uncertainty regarding their emissive properties. In particular, very few reports of the excitation spectra of eumelanin have appeared in the literature, and those that have appeared are neither consistent nor conclusive.

A summary of previous results is presented in Table 1. One of the earliest fluorescence studies examined excitation and emission spectra of eumelanin and pheomelanin, extracted from hair of different strains of mice.[8] The authors found that the excitation spectra of eumelanin and pheomelanin exhibited broad asymmetric bands with a great deal of fine structure and peaks at 376 and 308 nm, respectively. The emission wavelengths for these data were not specified. However, their extraction method involved acid/base treatments, which have been shown to disrupt cellular morphology[12] and may also severely alter the molecular structure.

| Source | Excitation Peaks | Emission Wavelength | Reference |
|---|---|---|---|
| Yellow mouse hair | 308 nm (4.03 eV) | Unspecified | 8 |
| Black mouse hair | 376 nm (3.30 eV) | Unspecified | 8 |
| Human retina | 450, 370 nm (2.76, 3.35 eV) | 570 nm (2.18 eV) | 9 |
| Bovine retina | 450, 370 nm (2.76, 3.35 eV) | 570 nm (2.18 eV) | 9 |
| Bovine retina | 470, 400, 420 nm (2.64, 3.10, 2.95 eV) | 540 nm (2.30 eV) | 10 |
| Synthetic eumelanin | 470 nm (2.64 eV) | 540 nm (2.30 eV) | 10 |
| Synthetic eumelanin | 370, 440 nm (3.35, 2.82 eV) | 540 nm (2.30 eV) | 11 |
| Synthetic eumelanin | 370, 420 nm (3.35, 2.95 eV) | 500 nm (2.48 eV) | 11 |
| Sepia ink eumelanin | 250, 390 nm (4.96, 3.18 eV) | 500 nm (2.48 eV) | 7 |

Table 1: Summary of previous excitation studies

    Boulton *et al.* studied human and bovine retinal eumelanin granules, finding that at an emission wavelength of 570 nm, excitation spectra of adult eumelanin granules for both species exhibited a broad, asymmetric band, peaking at 450 nm with a shoulder near 370 nm.[9] While the basic shape of the curve showed some similarity to that of eumelanin in the previous study, the peak was offset by 75 nm and the shoulder was much more pronounced. Another study of bovine retinal eumelanin (using an emission wavelength of 540 nm) also reported the excitation spectrum as an asymmetric band, but this time with an excitation peak at 470 nm and much more prominent shoulders at 400 and 420 nm.[10] Excitation spectra for synthetic eumelanin, published in the same report, did not exhibit the shoulders, appearing as a symmetric curve with a sharp peak at 470 nm. It is surprising that none of these spectra showed any similarity to the absorption spectrum of eumelanin. Such similarity is intuitively expected because greater absorption should lead to greater emission and hence, fluorescence intensity. This is a general property of organic systems.

    Two more recent studies examining the excitation spectra of eumelanins provided sharply contrasting data with the previous reports. One study examined excitation spectra for synthetic eumelanin dissolved in DMSO for several emission wavelengths between 460 and 700 nm.[11] Their spectra exhibited a peak near 370 nm (whose position was constant but whose intensity decreased with increasing emission wavelength) and a broader band that varied in position and intensity with increasing emission wavelength. A different study of eumelanin extracted from the ink of *Sepia officialis* (a good model for mammalian eumelanin) revealed spectra exhibiting two broad peaks centred at 250 nm and 390 nm for an emission wavelength of 500 nm.[7] In both studies, the spectral baseline decreases with increasing wavelength, although not with the same form as the absorption, as might be expected. These data contrast sharply with those of the earlier studies – both in the shape of the spectral baseline and in the locations of the peaks.

    Eumelanin has strong, broadband absorption (Fig. 1a) and only weak emission. Measurements of fluorescence are therefore strongly affected by attenuation of the incident beam (the inner filter effect) and emission reabsorption. Due to the non-linear shape of eumelanin absorption, these effects skew emission spectra in a non-trivial way, altering

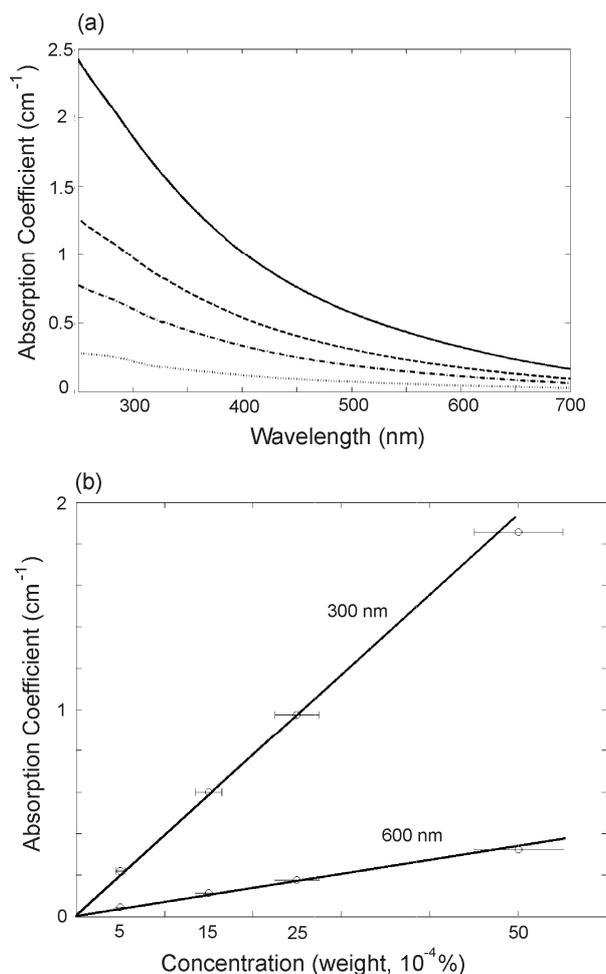

Figure 1: (a) Absorbance spectra of synthetic eumelanin at four concentrations: 0.0005% (dot), 0.0015% (dot-dash), 0.0025% (dash), and 0.005% (solid) by weight. (b) Absorption coefficients at 300 and 600 nm as functions of concentration for four synthetic eumelanin solutions. Concentration errors were estimated to be less than $1\times10^{-4}$% by weight.

both the spectral lineshape and the intensity.[13] Previous studies did not account for these properties, and their results will therefore vary in shape depending upon concentration and the exact shape of the absorbance spectrum (which varies for different melanin types). This is a possible explanation for much of the disagreement within the eumelanin spectroscopy literature.

In this study, we investigate the photophysical properties of eumelanin, as revealed by their excitation spectra, and their dependence on solute concentration and emission wavelength. A correction (previously shown to be highly effective for emission spectra[13, 14]) is applied for the first time to excitation spectra, in order to account for absorption effects and re-construct the excitation profiles. We present these excitation spectra as the most accurate and comprehensive to date. They are suitable for both direct qualitative comparision and quantitative analysis.

**Experimental Section**
Sample Preparation and Spectroscopy
Synthetic eumelanin (dopamelanin) derived by the nonenzymatic oxidation of tyrosine was purchased from Sigma Aldrich (Sydney, Australia) and underwent acid precipitation in order to remove small molecular weight components, following the method of Felix *et al*.[15] Briefly, dopamelanin (0.0020 g) was mixed in 40 mL high-purity 18.2 MΩ MilliQ deionized water and 0.5 M hydrochloric acid was added to bring the pH to 2. Solutions were centrifuged and the black precipitates were repeatedly washed in 0.01 M hydrochloric acid and then deionized water. Solutions of the yield were prepared at a range of concentrations (0.0005 – 0.0050%) by weight macromolecule in deionized water. To aid solubility, the pH of the solutions was adjusted to ~10 using 0.01 M NaOH (as in previous studies[13, 14]), and the solutions were gently stirred. Given that high pH also enhances polymerization,[16] this adjustment also ensured that the presence of any residual monomers or small oligomers in the solution was minimised. Under such conditions, pale brown, apparently homogeneous eumelanin dispersions were produced.

Absorption spectra between 250 and 700 nm were recorded for the synthetic eumelanin solutions using a Perkin Elmer (Melbourne, Australia) λ40 spectrometer. An increment of 2 nm, scan speed of 240 nm/min, and band pass of 3 nm were used. All spectra were collected using a 1 cm quartz cuvette. Solvent scans (obtained under identical conditions) were used for background correction. Fig. 1b shows the absorption coefficients at 300 and 600 nm as functions of concentration. The errors were estimated to be less than 1x10^-4% by weight, and the linear relationship is confirmed to be well within this error.

Photoluminescence excitation spectra were recorded for all concentrations using a Jobin Yvon (Paris, France) Fluoromax 3 fluorimeter. Excitation scans were recorded between 250 nm and the emission wavelength for emissions of 450, 490, 530, and 570 nm. A band pass of 3 nm and an integration time of 0.5 s were used. Again, solvent scans were acquired for background correction. The spectra were corrected for attenuation of the probe and reabsorption of the emission according to the procedure outlined below.

Correction Procedure

The appropriate correction for re-absorption and inner-filter effects has been shown[14] to be

$$I_{fl} = I_m \left( e^{\alpha_1 \left(a+\frac{w_1}{2}\right) + \alpha_2 \left(b+\frac{w_2}{2}\right)} \right) - I_{bg} \qquad (1)$$

where $I_{fl}$ is the actual emitted intensity, $I_m$ is the measured intensity (affected by re-absorption and probe beam attenuation), $I_{bg}$ is the background intensity, $\alpha_1$ and $\alpha_2$ are the absorption coefficients at the excitation and emission wavelengths respectively, and $a$, $b$, $w_1$ and $w_2$ are physical dimensions as shown in Fig. 2 (note that this is written in a different form from Ref. 14).

This correction was shown to be excellent as long as $\alpha_1(a+w_1/2) \ll 1$ and $\alpha_2(b+w_2/2) \ll 1$, which is true for dilute eumelanin samples. This report presents excitation spectra from more concentrated solutions where this approximation is no longer accurate. Therefore, we describe a more exact form of the correction, which also takes into account attenuation of the beam across the excitation volume, and use it to correct our spectra for attenuation effects.

In essence, we need to relate the measured fluorescence intensity $I_m$ to the "true" fluorescence intensity $I_{fl}$ that would be obtained with no attenuation or re-absorption effects. To derive this, we turn again to the experimental setup shown in Fig. 2 where the incident beam has a finite width, and the excitation occurs in an area $w_1 * w_2$, approximately 0.2 cm in each direction. The optical geometry of the fluorimeter makes this an excellent approximation. Note that everything is uniform in the $z$-direction and this dimension does not enter into the calculations. We define $I_0$ to be the total light intensity entering the cuvette (this has units of power rather than power per unit area, consistent with the "counts per second" measured by the detector).

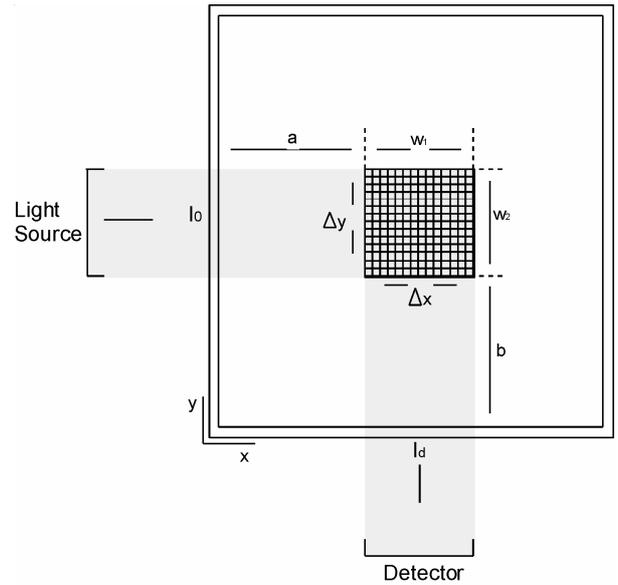

Figure 2: Diagram of cuvette. The excitation volume is assumed to be 0.2 cm in length and width. See text for details.

Consider first the theoretical case of no attenuation effects. In this case, the light intensity incident on the excitation volume is also $I_0$ (neglecting scattering) and the beam intensity leaving the excitation volume is given by the familiar Beer-Lambert law as

$$I = e^{-\alpha_1 w_1} I_0 \qquad (2)$$

The light absorbed over the width $w_1$ is then

$$\Delta I = I_0 - I_0 e^{-\alpha_1 w_1} \approx \alpha_1 w_1 I_0 \qquad (3)$$

where the exponential has been approximated with a first-order Taylor expansion. The resulting fluorescence received by the detector (not including any background signal) is given by

$$I_{fl} = \alpha_1 w_1 Q C I_0 \quad (4)$$

where $Q$ is a "quantum yield-like" factor, representing the ratio of photons emitted *at the measured wavelength only* to photons absorbed (at the excitation wavelength) and $C$ represents the fraction of emitted photons received by the detector. $C$ is dependent only on the geometry of the experimental setup and the detector sensitivity, whereas $Q$ is dependent on the excitation and emission wavelengths.

To derive the measured fluorescence intensity, we divide the excitation region into small subvolumes of size $\Delta x * \Delta y$, where we will ultimately let $\Delta x$ and $\Delta y$ approach zero. Dividing the incident light equally along the $y$-direction and accounting for probe attenuation, the incident intensity on the subvolume at position $(x,y)$ is given by

$$I(x,y) = e^{-\alpha_1 x} I_0 \frac{\Delta y}{w_2} \quad (5)$$

and the light intensity absorbed in that subvolume is

$$\Delta I = e^{-\alpha_1 x} I_0 \frac{\Delta y}{w_2} \left(1 - e^{-\alpha_1 \Delta x}\right) \approx \alpha_1 \Delta x\, e^{-\alpha_1 x} I_0 \frac{\Delta y}{w_2} \quad (6)$$

Note that this approximation becomes exact in the limit as $\Delta x$ approaches zero. Multiplying by $w_1/w_1$ for future convenience, the fluorescence reaching the detector from this sub-volume is

$$I_d(x,y) = \frac{\Delta x\, e^{-\alpha_1 x}}{w_1} (\alpha_1 w_1 Q I_0) \frac{\Delta y}{w_2} e^{-\alpha_2 y} \quad (7)$$

where $Q$ and $C$ are as above, and $e^{-\alpha_2 y}$ represents reabsorption of the emission within the cuvette. Summing over $x$ and $y$ yields the total fluorescence from the excitation volume received by the detector, $I_d$. Letting $\Delta x$ and $\Delta y$ approach zero, the sums become integrals:

$$I_d = \frac{\alpha_1 w_1 Q C I_0}{w_1 w_2} \int_a^{a+w_1} \int_b^{b+w_2} e^{-\alpha_1 x} e^{-\alpha_2 y} dy\, dx \quad (8)$$

Evaluation of the integrals yields

$$I_d = \frac{\alpha_1 w_1 Q C I_0}{w_1 w_2} \left\{ e^{-\alpha_1 a} e^{-\alpha_2 b} \frac{e^{-\alpha_1 w_1} - 1}{-\alpha_1} \frac{e^{-\alpha_2 w_2} - 1}{-\alpha_2} \right\} \quad (9)$$

Experimentally, the measured fluorescence $I_m$ also includes a background signal $I_{bg}$ due to Raman scattering in the solvent (plus impurity emission and dark noise, which are negligible). As this will also be proportional to incident intensity (proportionality factor $\beta$) and subject to re-absorption effects, the above procedure can be repeated and gives the same correction factor. Combining this with Eq. 4, the true fluorescence intensity $I_{fl}$, the fluorescence as expected in the absence of attenuation (from Eq. 4), can be expressed as a function of the measured fluorescence $I_m$:

$$I_{fl} = I_m e^{\alpha_1 a} e^{\alpha_2 b} \frac{\alpha_1 w_1}{1 - e^{-\alpha_1 w_1}} \frac{\alpha_2 w_2}{1 - e^{-\alpha_2 w_2}} - I_{bg} \quad (10)$$

This correction is calculated for each excitation and emission wavelength in our study by pointwise operations. Background scans were performed with deionized water under identical instrumental conditions. Note that in the limit where the absorption in the excitation region is small ($\alpha_1 w_1 << 1$, $\alpha_2 w_2 << 1$), $(1 - e^{\alpha w})$ can be approximated by $\alpha w$, restoring the expression in Eq. 1.

**Results and Discussion**

Uncorrected fluorescence excitation spectra of synthetic eumelanin for the different concentrations and emission wavelengths are shown in Fig. 3. The sharp peaks at 265 nm and 285 nm in the 530 and 570 nm emission plots, respectively, correspond to Rayleigh scattering bands, and two further features are of particular interest. First, there is a peak, most intense in the low concentration solutions, that shifts in position from 390 nm (450 nm emission) to 475 nm (570 nm emission). This is due to Raman scattering from the solvent, as is evident from background scans taken of the solvent without eumelanin (a second-order Raman band is also visible at 258 nm in the 570 nm emission plot). While the background intensity due to the solvent should be independent of fluorophore concentration, this is not found to be the case. Secondly, the fluorescence intensity should be directly proportional to the fluorophore concentration. While there is an increase in intensity with increasing concentration, the increase is much more pronounced at higher excitation wavelengths (where eumelanin

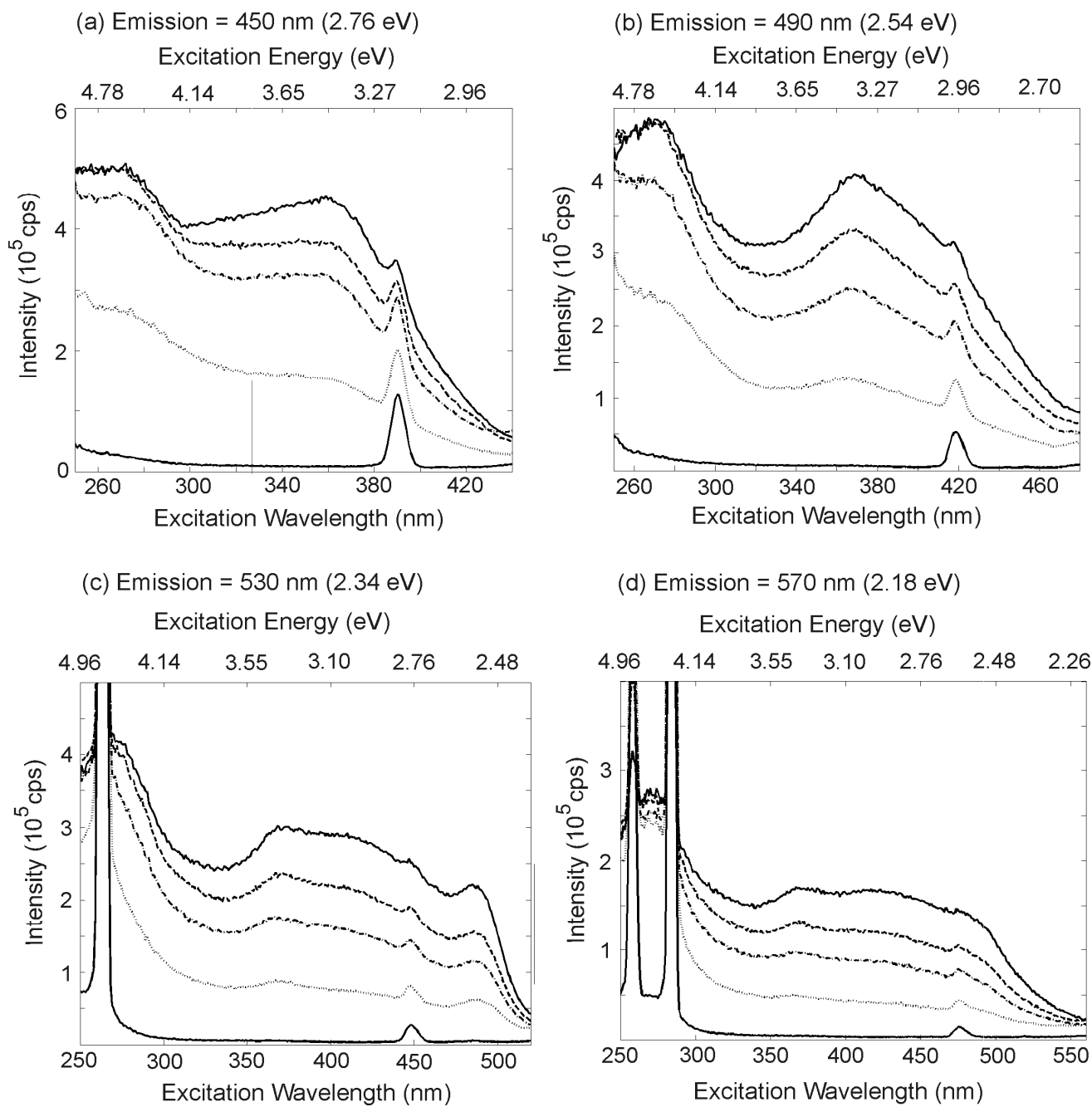

Figure 3. Uncorrected fluorescence excitation scans for four emission wavelengths (a: 450, b: 490, c: 530, d: 570 nm) and for four synthetic eumelanin concentrations: 0.0005% (dot), 0.0015% (dot-dash), 0.0025% (dash), and 0.0050% (solid) by weight. An excitation scan for the solvent (solid, bottom spectrum) is also shown.

absorbance is weakest), resulting in a change in spectral shape.

These effects can be explained by probe beam attenuation and re-absorption of emission. With increasing concentration, more light is absorbed, less light reaches the detector and measured intensities are reduced. Since the absorbance spectrum of eumelanin increases with increasing excitation energy, this effect is most prominent at short wavelengths. Attenuation and re-absorption also explain the very weak Raman signal at higher concentrations, as very little of the Raman scattered light is transmitted through the sample at higher concentrations. Clearly, re-absorption and probe beam attenuation severely affect eumelanin excitation and emission spectra.

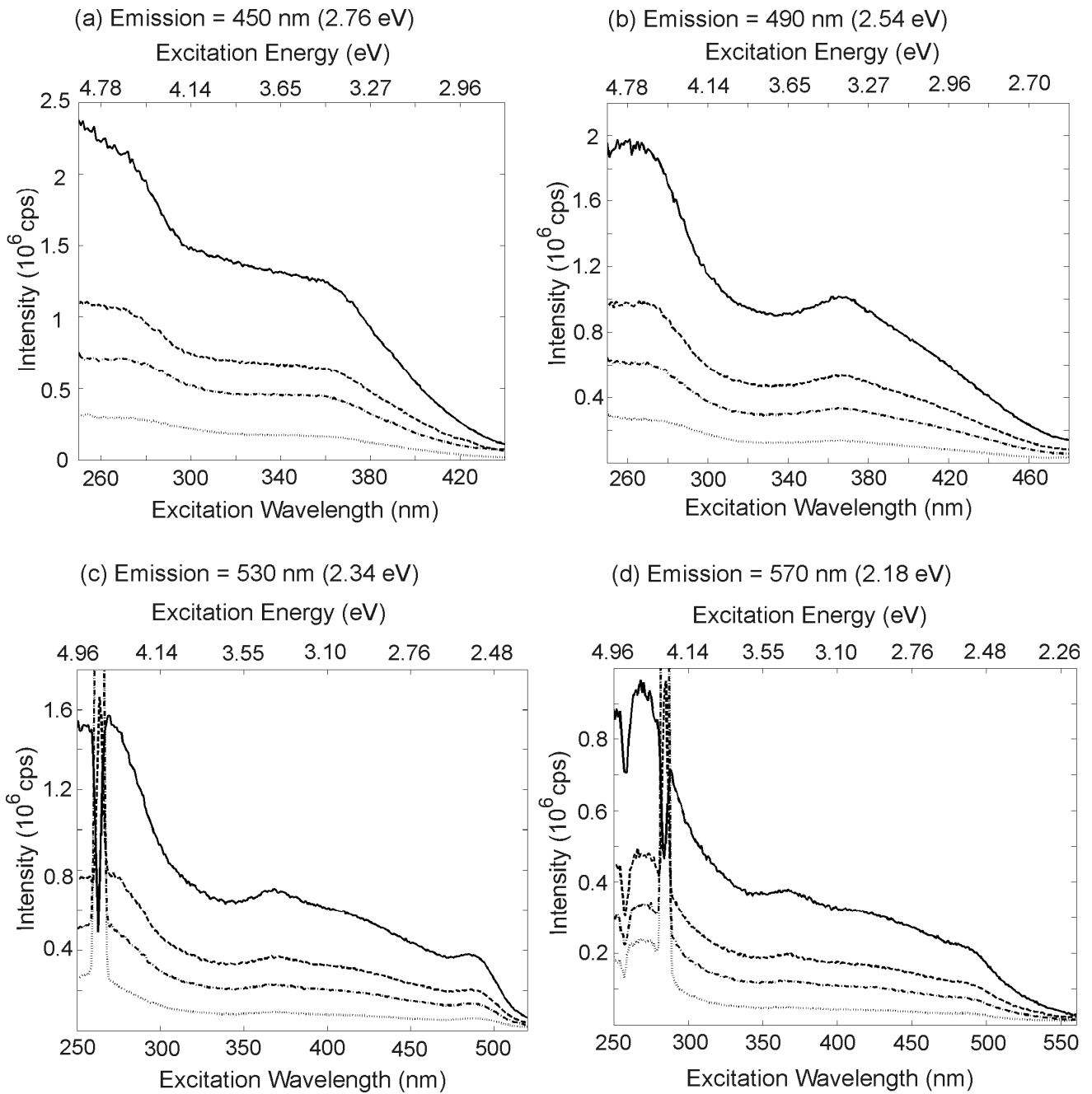

Figure 4: Corrected excitation scans for four emission wavelengths (a: 450, b: 490, c: 530, d: 570 nm) and for four synthetic eumelanin concentrations: 0.0005% (dot), 0.0015% (dot-dash), 0.0025% (dash), and 0.0050% (solid) by weight.

These effects also likely underlie many of the discrepancies among the previously published data. While some can be attributed to differences in the eumelanin source (e.g. Sepia ink vs. bovine retinal epithelium), the differences in peak location and spectral shape are more significant than would be expected from this alone. The data for Sepia eumelanin, which do show a decline in the baseline, most closely resemble the low-concentration data; whereas the retinal eumelanin data, which do not exhibit as much of a decline, are more similar to the high-concentration data. Thus, these differences may simply reflect artifacts in measurement rather than any true differences in fluorescence.

In order to remove these artifacts from our spectra, raw spectra were corrected according to the procedure described above. The correction in Eq. 1, applied to raw emission spectra of synthetic eumelanin, has been previously shown to restore the linear relationship between the emission peak intensity and eumelanin concentration and allowed for precise determination of the quantum yield, which was shown to be on the order of $10^{-4}$ and dependent on excitation wavelength.

Similarly, using Eq. 10, correction of the excitation spectra removed the Raman scattering peak entirely and restored the expected increase in fluorescence intensity with increased concentration (Fig. 4). Moreover, the corrected spectra assume a shape very similar to that of the absorption profile of eumelanin, as expected, since the increased absorption found at lower wavelengths should result in increased emission intensity. What is of further interest is that the apparent multiple band structure between 350 and 450 nm (most evident for 530 and 570 nm emission) disappears, leaving only a single peak at 365 nm (3.40 eV). This suggests that the feature at 420 nm does not correspond to real electronic transitions, but is rather an artifact resulting from the wavelength-dependent attenuation of the probe and emission. It is important to note that Eq. 1 and Eq. 10 produce virtually identical results except at high concentrations or low wavelength (high absorption coefficient).

The corrected excitation spectra in Fig. 4 are further validated by virtue of their close correspondence to emission spectra in Fig. 5. For instance, the emission spectrum for 350 nm excitation reaches a maximum at 450 nm, decreasing monotonically with increasing emission wavelength. This is also seen in Fig. 4 by noting the fluorescence intensity at 350 nm excitation in each plot; the maximum is found at 450 nm emission and decreases monotonically. Similarly, the emission spectrum for 440 nm excitation reaches a maximum near 510 nm, and following the intensity at 440 nm excitation through the four frames of Fig. 4 also reveals a maximum near 510 nm emission. Moreover, the intensity of this maximum can be shown to be less than half that of the 350 nm maximum, as is also evident in Fig. 5.

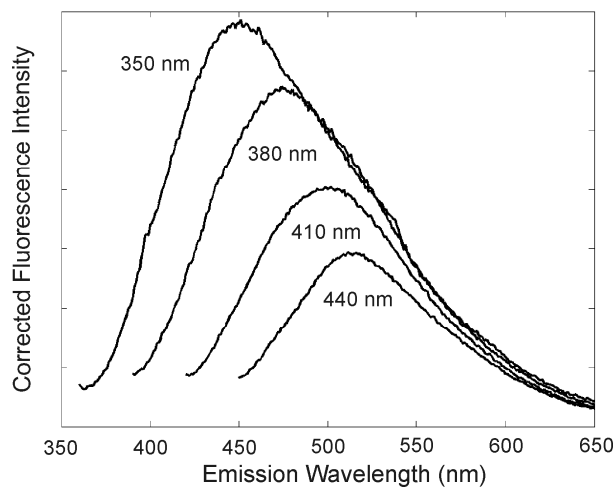

Figure 5: Corrected photoluminescence emission scans of 0.0025% eumelanin for four excitation wavelengths. Peak position and intensity are clearly wavelength-dependent.

Most significantly, the position of the peak at 365 nm (3.40 eV) is independent of emission wavelength. Corrected emission spectra exhibit a peak which is red-shifted as the excitation wavelength increases (Fig. 5). This has been speculated to be due to selective pumping of chemically distinct species within the eumelanin compound, each with different fundamental highest occupied molecular orbital (HOMO) – lowest unoccupied molecular orbital (LUMO) energy gaps.[14] This trend is not observed in the excitation spectra. Instead, a distinct excitation peak appears at 365 nm for all emission wavelengths studied.

Characterization of this peak is challenged by the limited data available on melanin subunits. Melanin has long been considered to be a large heteropolymer, with extensive cross-linking in a variety of positions.[17, 18] However, more recently, a new model for the secondary structure has emerged based upon the concept of stacked indolic oligomers. These oligomers are believed to consist of 4-7 monomers –most likely dihydroxyindole (DHI, also known as hydroquinone, HQ), dihydroxyindole-carboxylic acid (DHICA), indolequinone (IQ) and semiquinone (SQ).[5, 19, 20] The stacked oligomer model is supported by X-ray structural studies,[21-23] atomic force microscopy observations[5, 12] and numerous quantum chemical calculations at several levels of theory. In particular Bolivar-Marinez et al.[24] and Bochenek and Gudowska-Nowak[25] have employed the intermediate neglect of differential overlap (INDO) method and Il'ichev and Simon,[26] Powell et al.[27] and Stark et al.[28, 29] have used density functional theory (DFT) to study the energy structure of the individual monomers and small oligomers. The latter authors have combined *ab initio* with semi-empirical methods to study stacked systems and predict their absorption properties. In general, all of these computational studies have found the HOMO-LUMO gap of HQ to be approximately 4 eV, twice that of IQ and SQ, and the simulated absorption spectrum of HQ shows a strong maximum at 200 nm and a weaker band at 300 nm (4.14 eV). This is similar to the experimental data for HQ, but does not account for the observed secondary peak at 270 nm (4.60 eV). Simulations of larger oligomeric structures show that polymerisation leads to progressive red shifting of the gap and increased delocalisation of the electronic wavefunctions. Stark et al. in their latest paper have augmented these findings, demonstrating further red shifting with stacking. These data all lend support to the hypothesis that the broad band absorption and complex emission properties of melanins could be explained by invoking an ensemble model, where the system is composed of a number of chemical distinct species (monomers, oligomers, etc.), potentially but not necessarily stacked, and each with a different HOMO-LUMO gap in the UV, visible or near IR.

A more recent experimental study of the carboxylated form of DHI (DHICA) demonstrates a HOMO-LUMO gap of 3.8 eV for the monomer,[30] consistent with recent calculations.[31] Indeed, this is in good agreement with the observed peak at 365 nm (3.40 eV) in the present study, suggesting that this peak may reflect the presence of DHICA monomers in the solution. Since acid precipitation and the high final pH make the presence of monomers unlikely, it is conceivable that the DHICA monomer is singly linked to a larger structure, resulting in only a small redshift. Indeed, the increased intensity of the fluorescence peak suggests either that the chemical species responsible for the peak is more populous than other species (in which case a similar peak would appear in the absorbance spectrum) or that that species has a higher quantum yield. The latter hypothesis indicates weaker electron-phonon interaction, making the single-linking theory very plausible. One may even speculate that the DHICA monomer, because of its reduced reactivity, is the terminating unit for eumelanin secondary structures.

In plots of the relative differences in intensity between the corrected excitation spectra for higher concentrations and the corrected excitation spectra for 0.0005% eumelanin scaled by the concentration ratio (e.g. $(I_{0.0025\%} - 5*I_{0.0005\%})/(5*I_{0.0005\%})$), the 365 nm peak completely disappears from all spectra (Fig. 6). This indicates that this peak is a genuine feature of the eumelanin excitation spectrum, since its intensity is proportional to the concentration. These plots also demonstrate that while the correction in Eq. 10 restores the true excitation profiles remarkably, proportionality between corrected fluorescence intensity and concentration is not completely restored, as might be expected. If it were, then these differences would be zero at all wavelengths. Instead, we observe emissions below what would be expected. While it might be speculated that this is attributable to errors in solution concentration, this is not the case as absorbance spectra show very good linearity with concentration (Fig. 1b).

In Fig. 6, it is clear that (a) the relative differences are greatest at the higher wavelengths, and (b) the differences are only slightly dependent on concentration. The former is easily explained by the decreased signal intensity at high excitation wavelengths. The fact that the difference values are mostly constant around 20-30% with increasing concentration demonstrates the validity of the correction procedure even at high concentrations and suggests that the observed differences are due to scattering. While the loss in signal intensity is roughly proportional to concentration, so is the correction by the concentration ratio. The slight increase in absolute value, approaching an asymptote around 30%, might be attributable to multiple scattering events. The appearance of the solutions (light brown and not milky) and the form of the absorption profiles (closely fitting an exponential) would suggest that Mie scattering is negligible. Irrespective of the secondary structural model chosen (condensed nanoaggregates of oligomers or large extended heteropolymers), one would expect some Rayleigh scattering dependent upon the degree of molecular aggregation. Indeed, synthetic melanin aggregates have been shown to be as smaller than 5 nm in height[5] and Rayleigh scattering is valid for particle sizes less than 5% of the wavelength (13 nm for 260 nm excitation).

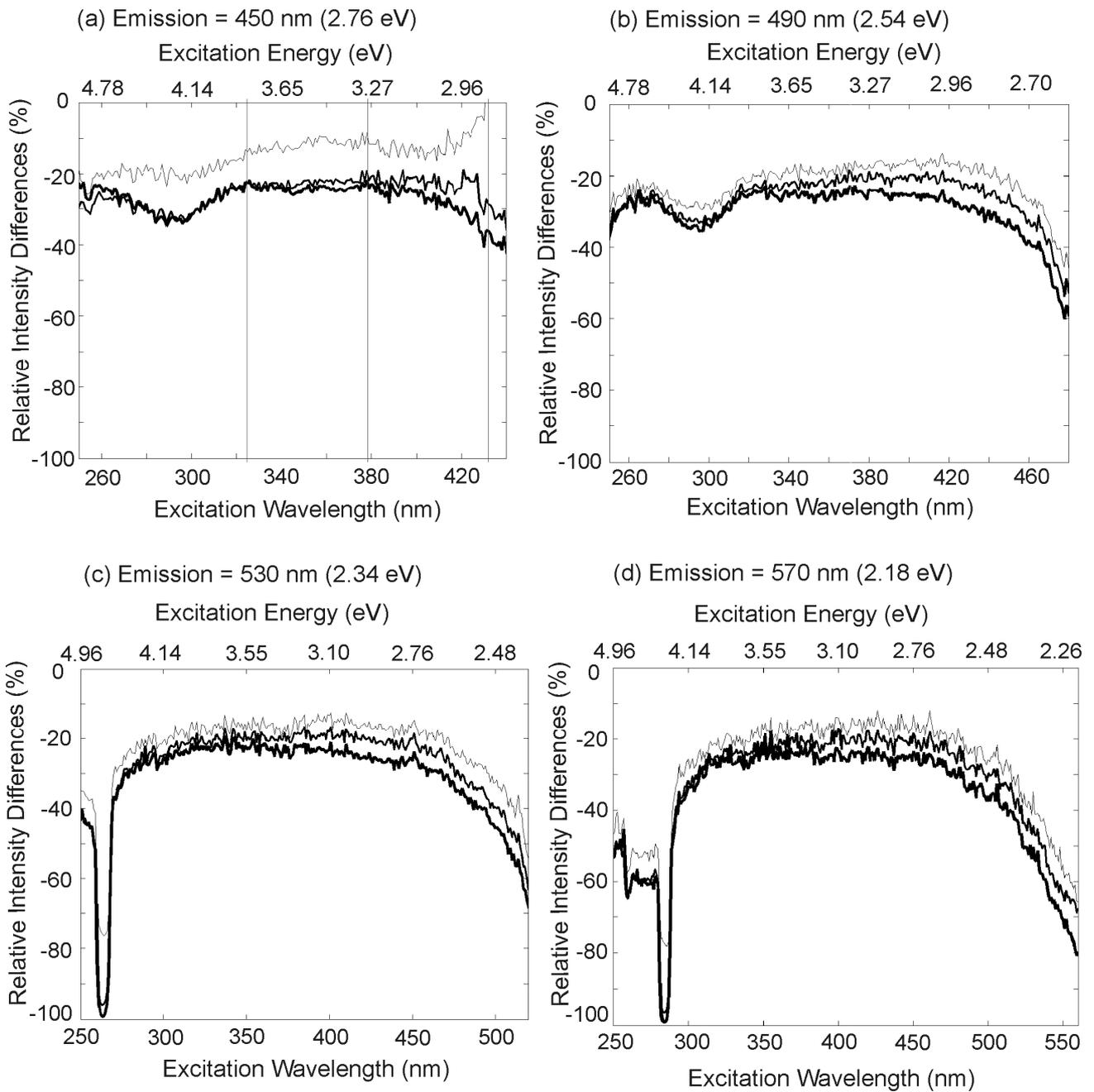

Figure 6: Differences in fluorescence intensity between the spectra and the low-concentration spectrum (0.0005%, scaled by factors of 3, 5, and 10) as a function of excitation wavelength for four emission wavelengths and for three concentrations: 0.0015% (thin), 0.0025% (medium), and 0.0050% (thick) by weight.

Other sources of error may be due to simplifications in the correction procedure. These include detected emission from scattered light outside the excitation volume, the approximation made in Eq. 3, and contributions to the background intensity that are independent of probe intensity or concentration (and are magnified by the correction procedure). The data presented here demonstrate the power of the correction procedure, and our corrected spectra exhibit the expected form and intensity values. We believe them to be the most accurate reported to date and useful for quantitative and qualitative analysis.

**Conclusions**
We have acquired fluorescence excitation spectra for synthetic eumelanin solutions at four different concentrations and using four different emission wavelengths between 450 and 570 nm. Comparison of these spectra to published excitation spectra suggest that differences among previous data may be due to artifacts resulting from probe attenuation and emission reabsorption. Application of a correction procedure to account for these factors demonstrates the expected similarity between the real excitation spectra and the absorption profiles of eumelanin. Moreover, the existence of a peak at 365 nm (3.40 eV), whose position is independent of emission wavelength, is clearly evident and suggestive of singly linked monomers in the stacked polymer structure. The correction method is observed retain validity even at high concentrations, although slight differences from expected values are observed, likely due to scattering effects. Given these caveats, we believe the spectra we have measured are suitable for direct qualitative comparison and full quantitative analysis. The peak at 365 nm can now be firmly asserted and needs to be a definitive feature in any structural model.

**References**
(1) Nicolaus, R. A., *Melanins*. Herman: Paris, 1968.
(2) Wakamatsu, K.; Fujikawa, K.; Zucca, F. A.; Zecca, L.; Ito, S. *J Neurochem* **2003,** *86*, 1015-1023.
(3) Hill, H. Z.; Hill, G. J. *Pigm Cell Res* **1987,** *1*, 163-170.
(4) Kipp, C.; Young, A. R. *Photochem Photobiol* **1999,** *70*, 191-198.
(5) Stark, K. B.; Gallas, J. M.; Zajac, G. W.; Golab, J. T.; Gidanian, S.; McIntire, T.; Farmer, P. J. *J Phys Chem B* **2005,** *109*, 1970-1977.
(6) Gallas, J. M.; Eisner, M. *Photochem Photobiol* **1987,** *45*, 595-600.
(7) Nofsinger, J. B.; Simon, J. D. *Photochem Photobiol* **2001,** *74*, 31-37.
(8) Ikejima, T.; Takeuchi, T. *Biochem Genet* **1978,** *16*, 673-679.
(9) Boulton, M.; Docchio, F.; Dayhawbarker, P.; Ramponi, R.; Cubeddu, R. *Vision Res* **1990,** *30*, 1291-1303.
(10) Kayatz, P.; Thumann, G.; Luther, T. T.; Jordan, J. F.; Bartz-Schmidt, K. U.; Esser, P. J.; Schraermeyer, U. *Invest Ophth Vis Sci* **2001,** *42*, 241-246.
(11) Teuchner, K.; Ehlert, J.; Freyer, W.; Leupold, D.; Altmeyer, P.; Stucker, M.; Hoffmann, K. *J Fluoresc* **2000,** *10*, 275-281.
(12) Liu, Y.; Simon, J. D. *Pigm Cell Res* **2003,** *16*, 606-618.
(13) Riesz, J.; Gilmore, J.; Meredith, P. *Spectrochim Acta A* **2005,** *61*, 2153-2160.
(14) Meredith, P.; Riesz, J. *Photochem Photobiol* **2004,** *79*, 211-216.

(15) Felix, C. C.; Hyde, J. S.; Sarna, T.; Sealy, R. C. *J Am Chem Soc* **1978,** *100*, 3922-3926.

(16) Adachi, K.; Wakamatsu, K.; Ito, S.; Miyamoto, N.; Kokubo, T.; Nishioka, T.; Hirata, T. *Pigm Cell Res* **2005,** *18*, 214-219.

(17) Galvao, D. S.; Caldas, M. J. *J Chem Phys* **1988,** *88*, 4088-4091.

(18) Pezzella, A.; dIschia, M.; Napolitano, A.; Palumbo, A.; Prota, G. *Tetrahedron* **1997,** *53*, 8281-8286.

(19) Clancy, C. M. R.; Simon, J. D. *Biochemistry-Us* **2001,** *40*, 13353-13360.

(20) Littrell, K. C.; Gallas, J. M.; Zajac, G. W.; Thiyagarajan, P. *Photochem Photobiol* **2003,** *77*, 115-120.

(21) Cheng, J.; Moss, S. C.; Eisner, M.; Zschack, P. *Pigm Cell Res* **1994,** *7*, 255-262.

(22) Cheng, J.; Moss, S. C.; Eisner, M. *Pigm Cell Res* **1994,** *7*, 263-273.

(23) Gallas, J. M.; Littrell, K. C.; Seifert, S.; Zajac, G. W.; Thiyagarajan, P. *Biophys J* **1999,** *77*, 1135-1142.

(24) Bolivar-Marinez, L. E.; Galvao, D. S.; Caldas, M. J. *J Phys Chem B* **1999,** *103*, 2993-3000.

(25) Bochenek, K.; Gudowska-Nowak, E. *Chem Phys Lett* **2003,** *373*, 532-538.

(26) Il'ichev, Y. V.; Simon, J. D. *J Phys Chem B* **2003,** *107*, 7162-7171.

(27) Powell, B. J.; Baruah, T.; Bernstein, N.; Brake, K.; McKenzie, R. H.; Meredith, P.; Pederson, M. R. *J Chem Phys* **2004,** *120*, 8608-8615.

(28) Stark, K. B.; Gallas, J. M.; Zajac, G. W.; Eisner, M.; Golab, J. T. *J Phys Chem B* **2003,** *107*, 3061-3067.

(29) Stark, K. B.; Gallas, J. M.; Zajac, G. W.; Eisner, M.; Golab, J. T. *J Phys Chem B* **2003,** *107*, 11558-11562.

(30) Tran, M. L.; Powell, B. J.; Meredith, P. *Submitted to Biophysical Journal* **2005**, Available at http://www.arxiv.org/ftp/q-bio/papers/0506/0506028.pdf.

(31) Powell, B. J. *Chem Phys Lett* **2005,** *402*, 111-115.